%% file: main-sig.tex
\documentclass[acmtog]{acmart}

\AtBeginDocument{%
  }

\renewcommand\footnotetextcopyrightpermission[1]{}

\input{macros}
\usepackage{amsmath}
\usepackage{booktabs}
\usepackage{multicol}
\usepackage{multirow}
\usepackage{siunitx}
\usepackage{hyperref}
\usepackage{xcolor}

\definecolor{pink}{rgb}{1.0, 0.6, 0.9} % Light pink color

\hypersetup{
    colorlinks=true,
    urlcolor=pink, % Sets the URL color to pink
}
\begin{document}

%\title{Feedforward Dynamic Concept Personalization via In Context Learning}
\title{Zero-Shot Dynamic Concept Personalization with Grid-Based LoRA}


\author{Rameen Abdal}
\affiliation{%
  \institution{Snap Research}
  % \city{New York}
  \country{Snap Inc.}
}
\author{Or Patashnik}
\affiliation{%
    \institution{Snap Research}
  % \city{New York}
  \country{Snap Inc.}
  }
\author{Ekaterina Deyneka}
\affiliation{%
   \institution{Snap Generative ML}
  % \city{New York}
  \country{Snap Inc.}
  }
\author{Hao Chen}
\affiliation{%
  \institution{Snap Generative ML}
  % \city{New York}
  \country{Snap Inc.}
  }
\author{Aliaksandr Siarohin}
\affiliation{%
   \institution{Snap Research}
  % \city{New York}
  \country{Snap Inc.}
  }

\author{Sergey Tulyakov}
\affiliation{%
  \institution{Snap Research}
  % \city{New York}
  \country{Snap Inc.}
  }

  \author{Daniel Cohen-Or}
\affiliation{%
  \institution{Snap Research}
  % \city{New York}
  \country{Snap Inc.}
  }

  \author{Kfir Aberman}
\affiliation{%
   \institution{Snap Research}
  % \city{New York}
  \country{Snap Inc.}
  }

\renewcommand{\shortauthors}{Abdal et al.}

% article.
\begin{abstract}
\input{0_abstract}
\end{abstract}

\begin{teaserfigure}
\centering
\centerline{\href{https://snap-research.github.io/zero-shot-dynamic-concepts/}}{\large \textcolor{pink}{\texttt{https://snap-research.github.io/zero-shot-dynamic-concepts/}}}
\includegraphics[width=\linewidth]{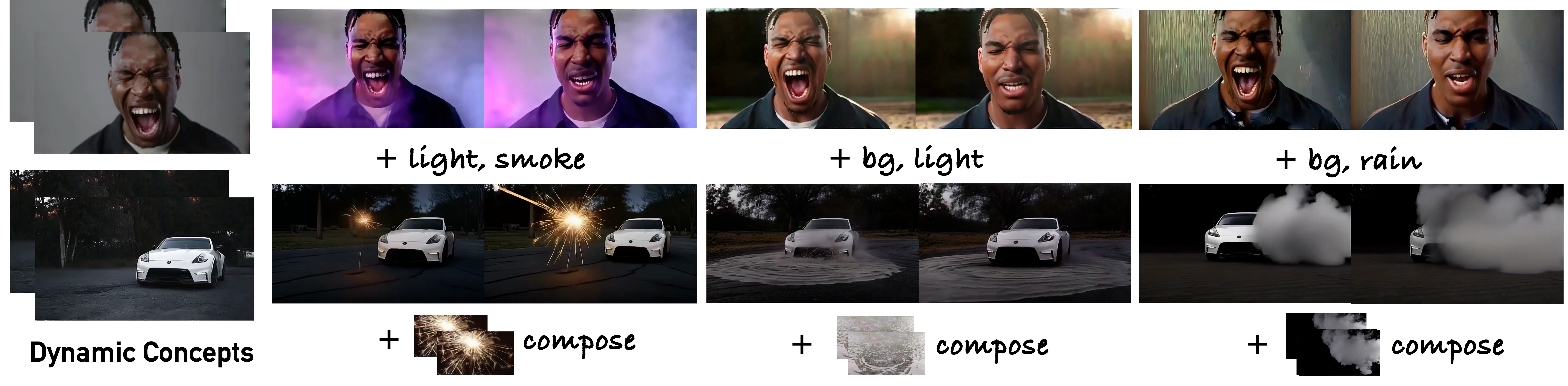}
\caption{We present a generalizable feedforward framework for personalizing text-to-video models with dynamic concepts. Our method enables text-driven editing (e.g., lighting changes, rain, smoke) and composition (e.g., sparks, water ripples, drifting smoke) without requiring per-video test-time fine-tuning.}
\label{fig:teaser}
\end{teaserfigure}



\maketitle

\input{1_introduction}

\input{2_related_work}

\input{3_method}

\input{4_experiments}

\input{5_conclusion}

\begin{acks}
We thank Gordon Guocheng Qian and Kuan-Chieh (Jackson) Wang for their feedback and support.
\end{acks}
\input{6_supplementary}


% \input{6_supplementary}
\bibliographystyle{ACM-Reference-Format}
\bibliography{sample-base}

\end{document}

%% file: macros.tex
\newif\ifdraft
\drafttrue
\ifdraft
\usepackage{xcolor}
\definecolor{newgreen}{rgb}{0.5, 0.6, 0.0}
\newcommand{\rac}[1]{{\color{teal}[\textbf{Rameen:} #1]}}
\newcommand{\opc}[1]{{\color{orange}[\textbf{Or:} #1]}}
\newcommand{\dcc}[1]{{\color{red}[\textbf{Danny:} #1]}}
\newcommand{\kac}[1]{{\color{purple}[\textbf{Kfir:} #1]}}
\newcommand{\willi}[1]{{\color{green}[\textbf{W} #1]}}

\newcommand{\todo}[1]{{\color{blue}[TODO: #1]}}

\else
\newcommand{\rac}[1]{}
\newcommand{\opc}[1]{}
\newcommand{\dcc}[1]{}
\newcommand{\kac}[1]{}
\newcommand{\willi}[1]{}
\newcommand{\todo}[1]{}

\fi

%% file: 0_abstract.tex
Recent advances in text-to-video generation have enabled high-quality synthesis from text and image prompts. While the personalization of dynamic concepts, which capture subject-specific appearance and motion from a single video, is now feasible, most existing methods require per-instance fine-tuning, limiting scalability. We introduce a fully zero-shot framework for dynamic concept personalization in text-to-video models. Our method leverages structured 2×2 video grids that spatially organize input and output pairs, enabling the training of lightweight Grid-LoRA adapters for editing and composition within these grids. At inference, a dedicated Grid Fill module completes partially observed layouts, producing temporally coherent and identity preserving outputs. Once trained, the entire system operates in a single forward pass, generalizing to previously unseen dynamic concepts without any test-time optimization. Extensive experiments demonstrate high-quality and consistent results across a wide range of subjects beyond trained concepts and editing scenarios.

%% file: 1_introduction.tex
\section{Introduction}

\begin{figure*}[t]
\begin{center}
\includegraphics[width=\linewidth]{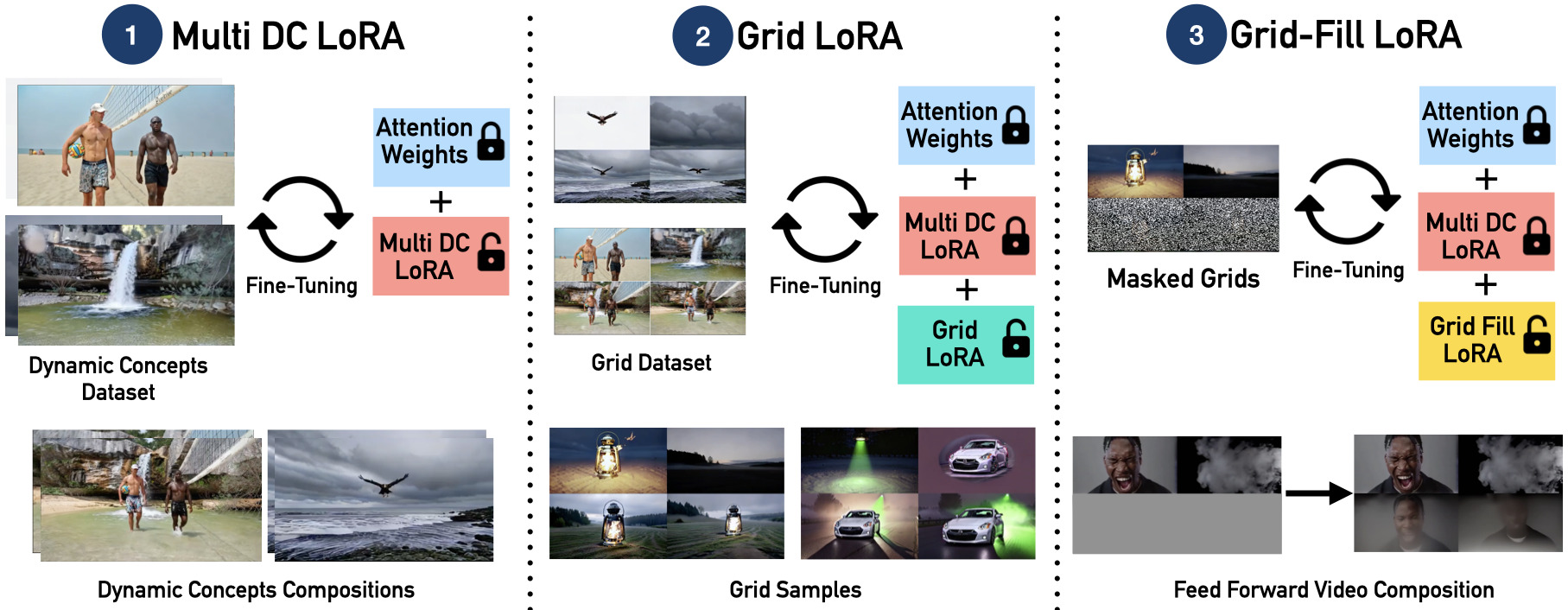}

\end{center}

\caption{Overview of our feedforward framework for dynamic concept personalization in text-to-video generation, eliminating the need for test-time fine-tuning. The system comprises three core modules: (i) Multi Dynamic Concept (DC) LoRA, which captures appearance and motion across diverse dynamic concepts from individual videos \cite{abdal2025dynamic} (ii) Grid LoRA, a layout-aware module trained on structured grids of these concepts; and (iii) Grid-Fill LoRA, which performs conditional inpainting to complete partially observed grids.}
\label{fig:method}
\end{figure*}

Recent advances in text-to-video generation have enabled the synthesis of high-resolution, temporally coherent clips from rich prompts, unlocking new opportunities in animation, content creation, and generative storytelling~\cite{ho2022imagen,blattmann2023stable,bar2024lumiere}. Despite this progress, personalized video generation remains a challenging task, particularly when it requires fine-grained control over subject-specific appearance and motion.

Current personalization approaches typically rely on per-instance fine-tuning, motion retargeting, or test-time optimization~\cite{wei2023dreamvideo,ren2024customize}. While these methods can yield strong results, they are computationally expensive, brittle to unseen inputs, and fundamentally limited in scalability.

A promising alternative has emerged in the form of dynamic concepts~\cite{abdal2025dynamic}, where small adapter modules (e.g., LoRA) are trained to capture both appearance and motion from a single input video. However, this approach still requires per-concept training, making it slow, non-scalable, and unsuitable for zero-shot personalization.

In this work, we introduce a fully feedforward framework for dynamic concept personalization in text-to-video diffusion models. Our method eliminates the need for per-instance optimization while enabling zero-shot generalization to new subjects and compositions.
Our key idea is to train layout-aware modules to generate structured 2×2 video grids, where each grid spatially arranges different variations of the same dynamic concept—through editing, composition, or both. By learning to produce temporally consistent and identity-preserving outputs across these grids, the model acquires strong spatial and contextual priors. We then introduce a Grid-Fill module that learns to complete partially filled grids, enabling targeted editing and composition in a single forward pass.

Concretely, our framework is composed of three core components:
(i) Multi Dynamic Concept (DC) LoRA: A unified adapter that encodes both appearance and motion from diverse single-video inputs.
(ii) Grid LoRA: A layout-aware module trained on structured 2×2 grids to support consistent spatial and temporal composition of dynamic concepts.
(iii) Grid-Fill LoRA: A conditional inpainting module that completes partially observed grids with identity-preserving, temporally coherent content.
Together, these components enable a novel grid-based training paradigm in which the model learns in-context transformations directly from spatially arranged input-output pairs. At inference time, our framework supports rich editing and composition in a single forward pass, without requiring any test-time adaptation.

We validate our approach through extensive experiments across a wide range of subjects and tasks. Results show that our method achieves high fidelity, strong temporal coherence, and broad generalization to unseen concepts, offering a scalable and practical solution for personalized, composable video generation.

%% file: 2_related_work.tex
\begin{figure*}[t]
\begin{center}
\includegraphics[width=\linewidth]{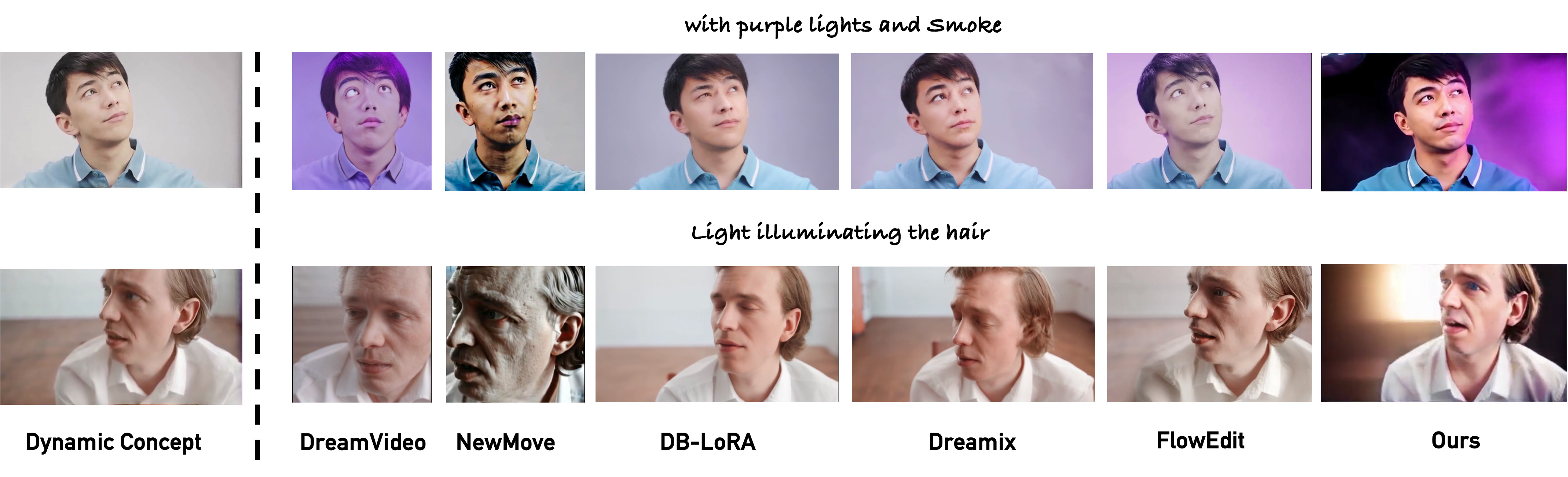}
\end{center}
\vspace{-0.5cm}
\caption{\textbf{Comparison with baselines.} Comparison of our method with baseline approaches (NewMove~\cite{materzynska2024newmove}, DreamVideo~\cite{wei2023dreamvideo}, DB-LoRA~\cite{simo,ruiz2023dreambooth}, and DreamMix~\cite{molad2023dreamixvideodiffusionmodels}) and FlowEdit~\cite{kulikov2024flowedit} on two editing scenarios: adding lights and smoke, and adding light that illuminates the hair. Our method demonstrates the ability to adhere to the prompt, while generating interactions that preserve subject appearance and motion.}

\label{fig:comparison}
\vspace{-0.3cm}
\end{figure*}

\begin{figure}[t]
\begin{center}
\includegraphics[width=1.0\linewidth]{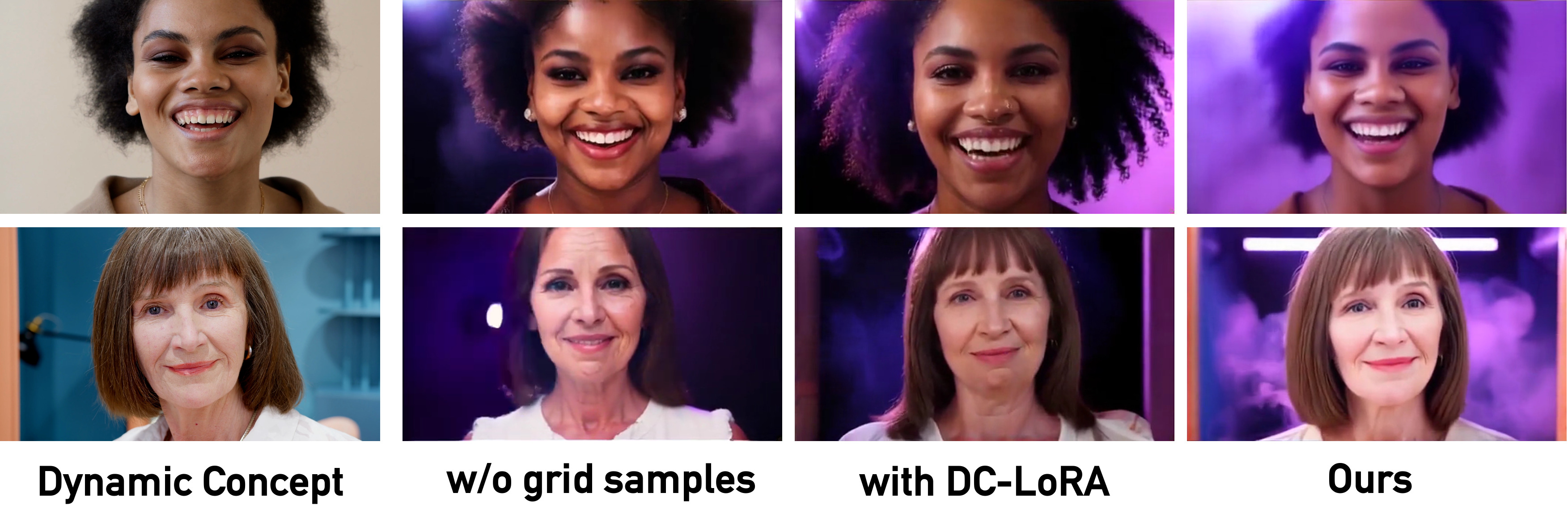}
\end{center}

\caption{\textbf{Ablation of Design Choices.} We ablate against two design choices i.e. augmenting the training set with Grid LoRA generated samples and removing the Multi-DC LoRA during inference.}
\label{fig:ablation}
\vspace{-0.2cm}

\end{figure}

\section{Related Work}

% \dcc{In the following I think I would distinguish between methods that are based on 2D foundation models like TokenFlow, and those that are build upon video foundation models.}

\paragraph{Text-to-Video Generation.}
 
Early work in text-to-video generation built on image diffusion models by introducing temporal layers to capture motion across frames. Models such as ImagenVideo~\cite{ho2022imagen} and Stable Video Diffusion~\cite{blattmann2023stable} extended U-Net backbones with temporal attention~\cite{hong2022cogvideo,singer2022make,guo2023animatediff}, enabling short video synthesis but struggling with scalability and temporal coherence. More recent approaches adopt fully transformer-based architectures with joint spatiotemporal attention, including Sora~\cite{sora_model}, RIN~\cite{RIN}, and SnapVideo~\cite{menapace2024snap}, culminating in large-scale systems like MovieGen~\cite{polyak2024movie}, HunyuanVideo~\cite{HunyuanVideo}, and CogVideoX~\cite{yang2024cogvideox}. While these models achieve high fidelity, they are designed for general video synthesis and offer limited support for user-controllable personalization or multi-concept composition.

\paragraph{Video Personalization.}
Video personalization aims to adapt a generative model to internalize the identity, appearance, or motion of a specific subject, enabling reuse across diverse scenes and prompts. In the image domain, personalization has seen rapid progress~\cite{cai2024dsd} with methods like DreamBooth~\cite{ruiz2023dreambooth}and IC-LoRA~\cite{lhhuang2024iclora}, which fine-tune or adapt pretrained diffusion models using a few examples. Extending these ideas to video requires maintaining temporal consistency in addition to spatial identity. Recent approaches address this by introducing LoRA modules into the temporal layers of video diffusion backbones~\cite{wu2024customttt,he2024idanimatorzeroshotidentitypreservinghuman}, or by learning personalized motion embeddings from reference videos~\cite{wei2023dreamvideo,materzynska2024newmove,zhao2023motiondirector}. Dynamic Concepts~\cite{abdal2025dynamic} represents a key step forward by enabling LoRA-based concept learning from a single video, capturing both appearance and motion. However, it still requires training a dedicated module per concept and lacks support for feedforward generalization. Our method builds on this foundation by introducing a structured, grid-based LoRA system that enables efficient, reusable, and composable dynamic concept personalization in a fully feedforward manner.

\paragraph{Video Editing.}
Video editing focuses on modifying an existing video to reflect a user-specified change (e.g., altering identity, adding motion, or changing style) while preserving other attributes such as structure or coherence. Many methods address this via test-time optimization or inversion, such as Customize-a-Video~\cite{ren2024customize}, Fate/Zero~\cite{qi2023fatezero}, DreamMix~\cite{molad2023dreamixvideodiffusionmodels}, and Tune-A-Video~\cite{wu2023tuneavideooneshottuningimage}. Others improve consistency through architectural design, using attention-based tracking or flow-guided propagation, as seen in TokenFlow~\cite{tokenflow2023}, RAVE~\cite{kara2024rave}, and FlowVid~\cite{liang2023flowvid}. Another set of works~\cite{kulikov2024flowedit, uniedit} , predominantly developed for the image domain and extended to videos, demonstrate local edits that align with the pixel but lack the ability to add interactions, camera perspectives, and compositions and hence are not inherently personalized. While effective in specific tasks, these approaches are often computationally expensive, task-specific, or limited in their ability to compose edits across multiple scenes or concepts. In contrast, our feedforward framework enables not only consistent identity preservation but also structured inpainting and multi-concept fusion, without requiring per-video optimization.

\paragraph{In-Context Generation}  
In-context generation lets a model adopt new tasks (such as segmentation, story generation, etc.) or concepts with a small set of example pairs.  In vision, Visual Prompting~\cite{visprompt} arranges input–output examples and a query into a grid and formulates the task as inpainting the masked cell.  Similarly, IC-LoRA~\cite{lhhuang2024iclora} uses the image as grids to learn to generate consistent characters using LoRA.  Building on these ideas, we employ grid-structured prompts and grids generated using Dynamic Concepts LoRA~\cite{abdal2025dynamic} to achieve consistent multi-pane editing or composition in feed forward fashion.

\paragraph{Composing Dynamic Concepts.}
Video composition—blending multiple dynamic elements into a coherent scene remains a difficult problem. Early works like Blended Diffusion~\cite{avrahami2022blended} and Break-A-Scene~\cite{avrahami2023break} introduced object blending and inpainting in the image domain, using cross-attention and region masks. 

Recent works extend these ideas to video by animating composed images~\cite{blattmann2023stable,chen2024videoalchemy,haCohen2024ltxvideo}, but often decouple appearance from motion, leading to artifacts and inconsistent dynamics. Another approach is to personalize image-based models using ~\cite{qian2024omniidholisticidentityrepresentation,wang2024moa}, and then animating the images using Image2Video frameworks~\cite{wan2025}. In contrast, we propose a unified framework that embeds dynamic concepts into a spatiotemporal LoRA representation, enabling lightweight, task-specific adapters to perform grid-based synthesis, inpainting, and composition, all in a single feedforward pass.

%% file: 3_method.tex
\begin{figure*}[t]
\begin{center}
\includegraphics[width=\linewidth]{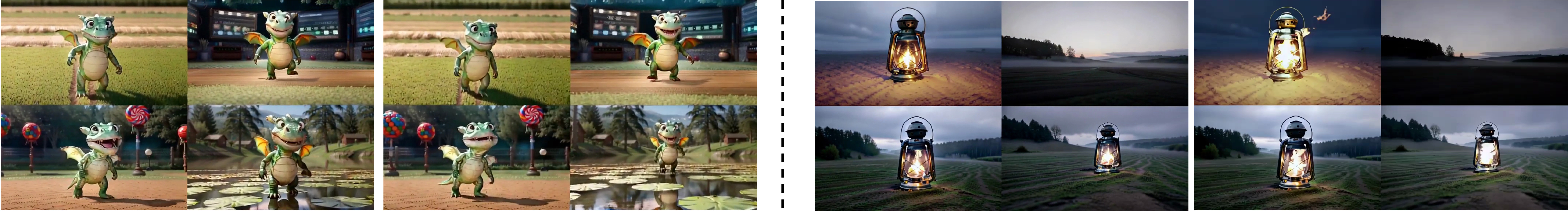}
\end{center}
\caption{\textbf{Grid LoRA Sampling.} Left: Sample from the Grid LoRA trained for consistent dynamic concepts. Right: Sample from the Grid LoRA trained for composition.}
\label{fig:sampling}
\end{figure*}

\begin{figure*}[t]
\begin{center}
\includegraphics[width=1.0\linewidth]{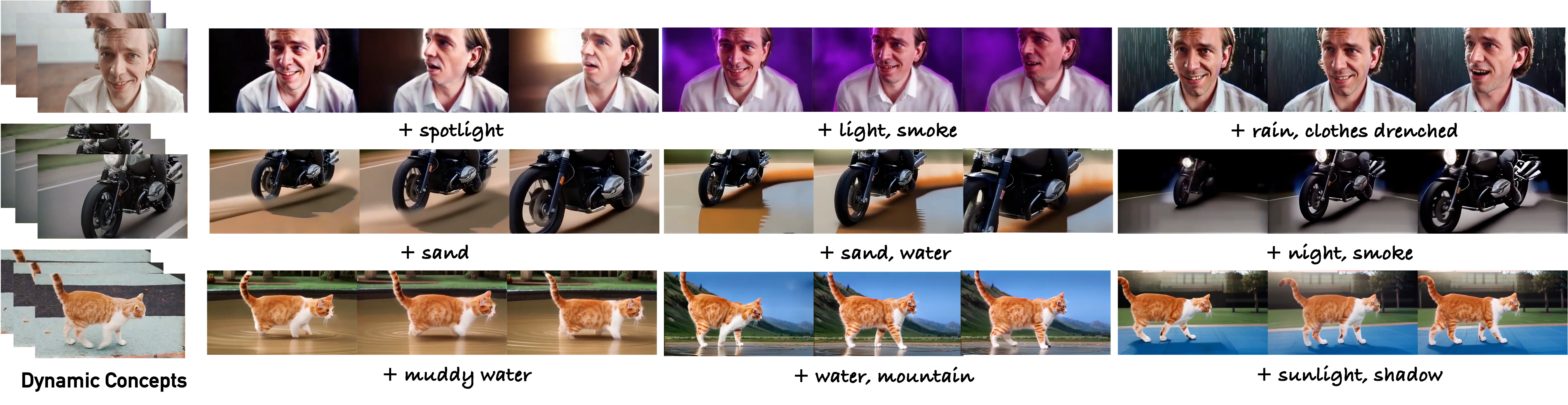}
\end{center}
\vspace{-0.3cm}
\caption{\textbf{Editing and Interactions.} Our feedforward Grid-Fill LoRA enables text-driven edits that go beyond pixel-aligned changes, supporting subject-aware personalization. Here the examples capture interactions while preserving appearance and motion. }
\label{fig:editing}
\vspace{-0.2cm}
\end{figure*}
\section{Method}

We present a feedforward framework for dynamic concept personalization using text-to video generative models, while eliminating the need for per-instance optimization or test-time fine-tuning \cite{abdal2025dynamic}. Our approach consists of three key modules:  
(i) \textbf{Multi Dynamic Concept (DC) LoRA} – a unified module for capturing the appearance and motion of various dynamic concepts from individual videos,  
(ii) \textbf{Grid LoRA} – a layout-aware module fine-tuned on structured grids of dynamic concepts, and  
(iii) \textbf{Grid-Fill LoRA} – a conditional inpainting module for completing partially observed grids. Together, these components enable dynamic concepts personalization for editing and composition in a fully feedforward manner. The overall architecture is illustrated in Fig.~\ref{fig:method}.

\subsection{Multi Dynamic Concept LoRA}

We begin by collecting a dataset of dynamic concepts, each represented by a subject performing an action or a concept in motion e.g. waterfall across frames of a video. We collect 11 of such videos (5 identities and 6 scenes). Following the set-and-sequence training strategy of Abdal et al.~\cite{abdal2025dynamic}, we fine-tune a LoRA module to capture both the appearance and motion characteristics of these dynamic subjects. Unlike the one-LoRA-per-subject convention of~\cite{abdal2025dynamic}, we consolidate all dynamic concepts into a single \textbf{Multi-DC LoRA}, enabling the model to generalize across subjects and generate coherent compositions from a unified parameter set. We use unique identifiers such as [person\_identity] and [action\_motion] to encode multiple concepts in a common weight space.

Concretely, for a base weight matrix \(W \in \mathbb{R}^{m \times n}\), we use two low-rank updates:
\[
\Delta W_{\mathrm{app}} = A_1B_1,\quad \Delta W_{\mathrm{mot}} = A_1B_2,
\]
where \(A_1 \in \mathbb{R}^{m \times r}\), \(B_1, B_2 \in \mathbb{R}^{r \times n}\), and \(r \ll \min(m,n)\). These represent static appearance (\(A_1B_1\)) and motion-specific (\(A_1B_2\)) variations, resulting in a final updated weight:
\[
W' = W + A_1B_1 + A_1B_2.
\]

We optimize Multi-DC LoRA using a flow-matching loss:
\[
\mathcal{L}_{\mathrm{flow}} = \mathbb{E}_{x_t, t} \left\| v_\theta(x_t, t) - \frac{\partial x_t}{\partial t} \right\|_2^2,
\] conditioned on appearance and motion tokens. Once trained, Multi-DC LoRA is frozen and reused as the base generator for all subsequent modules.

\subsection{Grid LoRA}

To learn compositionality and cross-entry consistency of dynamic concepts, we introduce \textbf{Grid LoRA}, a second-stage module finetuned on structured 2$\times$2 video grids. The method is not specific to this layout, one chose to have different layouts for this task e.g., 1$\times$3 in case of vertical videos stacked horizontally. Grid LoRA operates atop the frozen Multi-DC LoRA and is guided by text prompts that describe each cell. While Multi-DC LoRA maintains subject-specific visual features, Grid LoRA learns layout-aware interactions and compositional reasoning. During inference, reducing the effect of Multi-DC LoRA (reducing it influence or completely dropping it) enables the generation of novel combinations beyond the training set.

\paragraph{Composition}

We train Grid LoRA on structured compositions where the top row of the grids in the dataset contains distinct dynamic concepts (A and B), and the bottom row reflects a fusion of the two. Let \(x_v^{(C)} \in \mathbb{R}^{N \times D}\) denote spatio-temporal patch embeddings for video \(C\). We compute attention queries via:
\begin{align}
q_A &= (W_q + \Delta W_{q,A})\,x_v^{(A)}, \\
q_B &= (W_q + \Delta W_{q,B})\,x_v^{(B)}, \\
q_{\mathrm{Out}} &= \left(W_q + \tfrac{1}{2}(\Delta W_{q,A} + \Delta W_{q,B})\right)\,x_v^{(\mathrm{Out})}.
\end{align}

We enforce structured attention masking as follows:
- \(q_A\) attends to \(T_g \cup T_A\),
- \(q_B\) attends to \(T_g \cup T_B\),
- \(q_{\mathrm{Out}}\) attends to \(T_g \cup T_A \cup T_B \cup T_{\mathrm{Out}}\).

Text prompts are aligned with this structure: \texttt{[TOP\_LEFT]} and \texttt{[TOP\_RIGHT]} use concept-specific tokens \(T_A\) and \(T_B\), while the bottom row uses \(T_{\mathrm{Out}}\) (prompts corresponding to \texttt{[TOP\_LEFT]} and \texttt{[TOP\_RIGHT]}) to describe the fusion. All entries share global tokens \(T_g\), ensuring high-level semantic consistency.

\paragraph{Consistent Dynamic Concepts}

For cases where we wish to preserve appearance and motion across grid cells (e.g., cloning a concept across all four cells), we constrain the training by setting \(\Delta W_{q,A} = \Delta W_{q,B}\) and omitting the attention masking. This results in a single Grid LoRA adapter fine-tuned for consistency across entries.

\paragraph{Sampling Grids}

At inference, we employ structured prompts with similar layout as the training prompts to sample 2$\times$2 grids by applying only Grid LoRA atop the base generator (i.e. attenuating the Mult-DC LoRA). These outputs demonstrate high spatial-temporal consistency, even for unseen concept configurations as illustrated by Fig.~\ref{fig:editing} (second and third rows showing out-of-domain generalization beyond human subjects) and Fig.~\ref{fig:composition}.

\subsection{Grid-Fill LoRA}

To enable personalized grid synthesis from partial inputs, we introduce \textit{Grid-Fill LoRA}, a third-stage lightweight module trained specifically to inpaint missing cells in structured 2$\times$2 grids. Unlike traditional autoregressive or iterative generation approaches, Grid-Fill LoRA performs \textbf{non-autoregressive}, feedforward inpainting in a \textit{single pass}, enabling efficient and consistent generation across the entire grid. 
During training, we randomly mask one or more cells from a 2$\times$2 grid composed of different spatial layouts, motions, or environments, and optimize the module to reconstruct the missing parts while preserving coherence with the visible cells. This process is conditioned on the frozen Multi-DC LoRA weights (learned in previous personalization stages), ensuring that appearance and motion remain consistent with the underlying dynamic concept. At test time, this setup allows injection of a real video snippet (e.g., the top-left cell) into the grid as a fixed conditioning input. The Grid-Fill LoRA then fills in the remaining cells such that the synthesized completions remain personalized, temporally coherent, and semantically aligned with the input video. We define the training objective as a flow-matching reconstruction loss over the masked regions:
\[
\mathcal{L}_{\mathrm{grid\text{-}fill}} = \mathbb{E}_{x_t, t, M} \left\| v_\theta(x_t \odot M, t; T, W_{\mathrm{Multi\text{-}DC}}) - \frac{\partial x_t}{\partial t} \odot M \right\|_2^2,
\]
where $\mathbf{x}_t$ is the noisy input grid, $M$ is a binary mask denoting missing cells, $\mathbf{T}$ are the prompt tokens, and $\mathbf{W}_{\text{Multi-DC}}$ are the frozen LoRA weights from the earlier personalization stages. The loss encourages accurate flow prediction for the masked regions while keeping the unmasked cells fixed, enabling reliable completions in downstream applications. This approach supports creative synthesis and editing tasks where real and generated content can coexist in structured video layouts while maintaining high fidelity and identity coherence across the grid.

\subsection{Inference}

At test time, our inference procedure is:
\begin{enumerate}
    \item Sample or input a real dynamic concept to populate one or two grid cells depending on eiting or composition mode.
    \item Use Grid-Fill LoRA to complete the full 2$\times$2 grid in a single feedforward pass.
    \item Upsample the outputs to the desired video resolution. We use a light-weight convolution-based upsampler to achieve this.
\end{enumerate}

This pipeline enables high-fidelity, consistent, and personalized video generation without any per-instance optimization or fine-tuning.

%% file: 4_experiments.tex
\section{Experiment Settings}

\subsection{Evaluation Dataset}  
Following Dynamic Concepts~\cite{abdal2025dynamic}, we evaluate our framework on a curated human‐centric video dataset collected from \href{https://www.pexels.com/}{Pexels}. It comprises five distinct identities performing a mix of intricate as well as pronounced actions to test the limits of the methods. Each clip is 5 s at 24 FPS and 576×1024 resolution. This setup tests ability of our method to preserve motion coherence and achieve high‐quality personalization. Since our method facilitates zero-shot personalization of the videos, we consider edits such as "adding smoke and lighting", "adding rain", and "adding lighting from behind" to test the interaction of the dynamic concepts with these effects. To the best of our knowledge since there are no feedforward video personalization methods, we test our method with other methods on the editing task.

\subsection{Baselines}  
We benchmark against a diverse set of competitive approaches, covering UNet-based, feed-forward, architecture-agnostic, and LoRA-based methods. For UNet-based video personalization, we include DreamVideo~\cite{wei2023dreamvideo} and NewMove~\cite{materzynska2024newmove}. We also compare qualitatively to video version of FlowEdit~\cite{kulikov2024flowedit}, a recently published feed-forward editing method driven by text prompts (ArXiv, 2024), using the publicly available WAN~\cite{wan2025} model. To assess mixed image–video tuning, we adapt the UNet-based DreamMix~\cite{molad2023dreamixvideodiffusionmodels} into our DiT backbone. For architecture-agnostic personalization, we compare against DreamBooth LoRA~\cite{ruiz2023dreambooth,simo} as a representative per-instance LoRA baseline. Set-and-Sequence~\cite{abdal2025dynamic} shows lower performance of Textual Inversion~\cite{textual_inversion} in video personalization tasks so we omit this method for comaparison here. DreamMix and DreamBooth LoRA often fail to faithfully capture motion patterns either underfitting and losing identity fidelity or overfitting and introducing artifacts resulting in poor trade-offs between identity preservation and edit fidelity.
\begin{figure}[t]
\begin{center}
\includegraphics[width=1.0\linewidth]{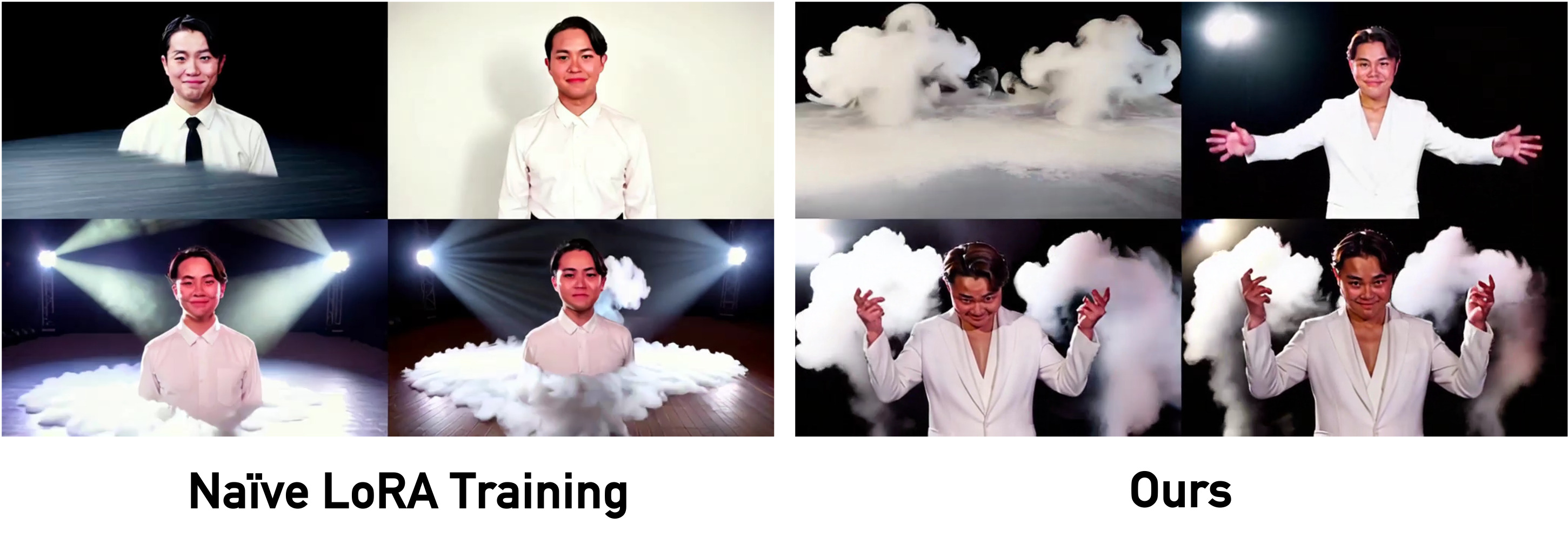}
\end{center}
\vspace{-0.5cm}
\caption{\textbf{Concept leaking.} Left: Naive finetuning of LoRAs result in concept leaking where the inputs (top row) leaks the information of the identity. Right: using structured LoRAs and attention masking mitigates this effect. }
\label{fig:leaking}

\vspace{-0.5cm}

\end{figure}

\begin{figure}[t]
\begin{center}
\includegraphics[width=1.0\linewidth]{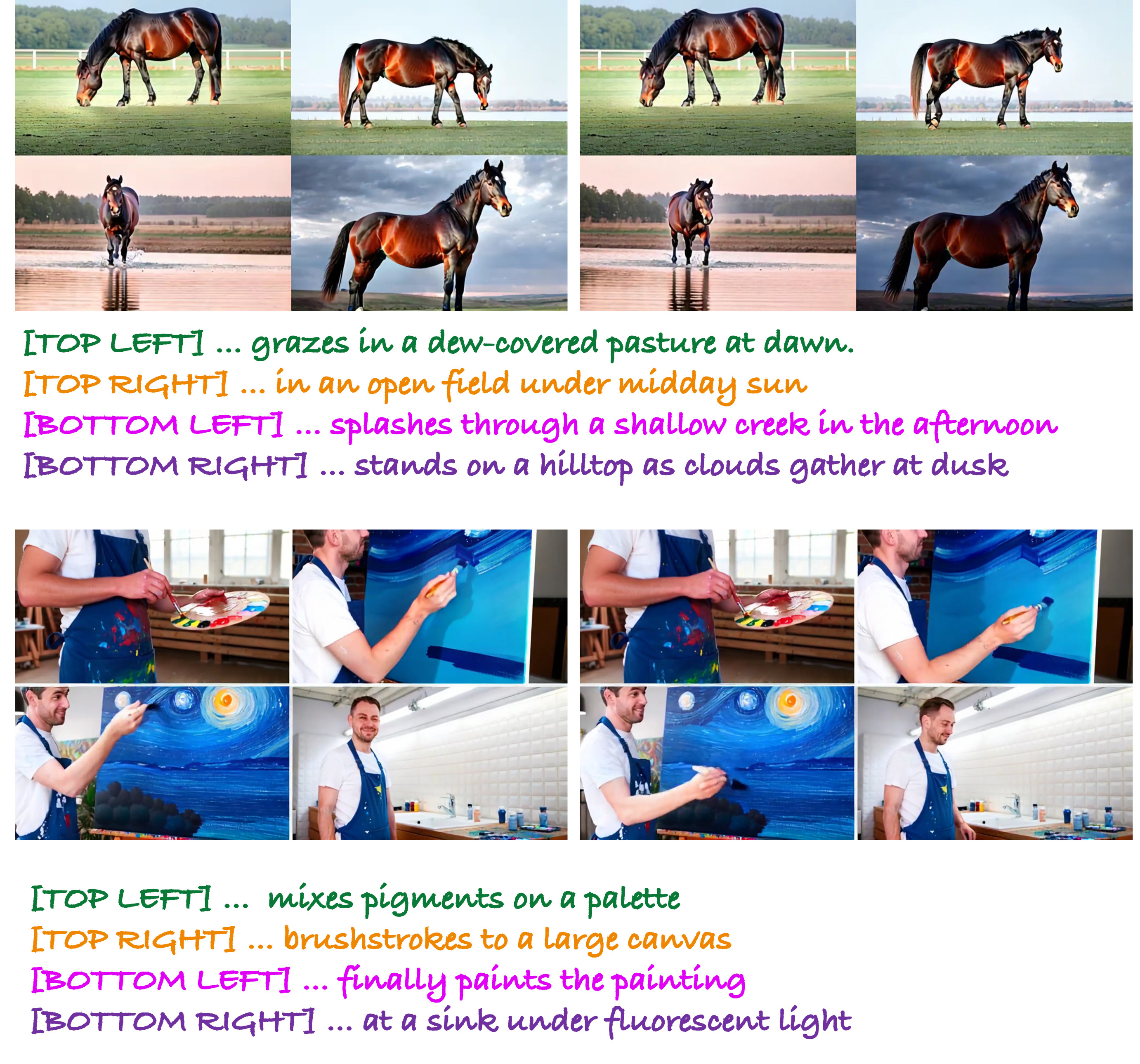}
\end{center}
\vspace{-0.5cm}
\caption{\textbf{Story Generation.} In the Grid LoRA setting for consistent dynamic concepts, we preserve identity and appearance while allowing variations in motion, interactions, and context enabling coherent story progression across different scenes.}
\label{fig:story}
\end{figure}

\begin{figure*}[t]
\begin{center}
\includegraphics[width=1.0\linewidth]{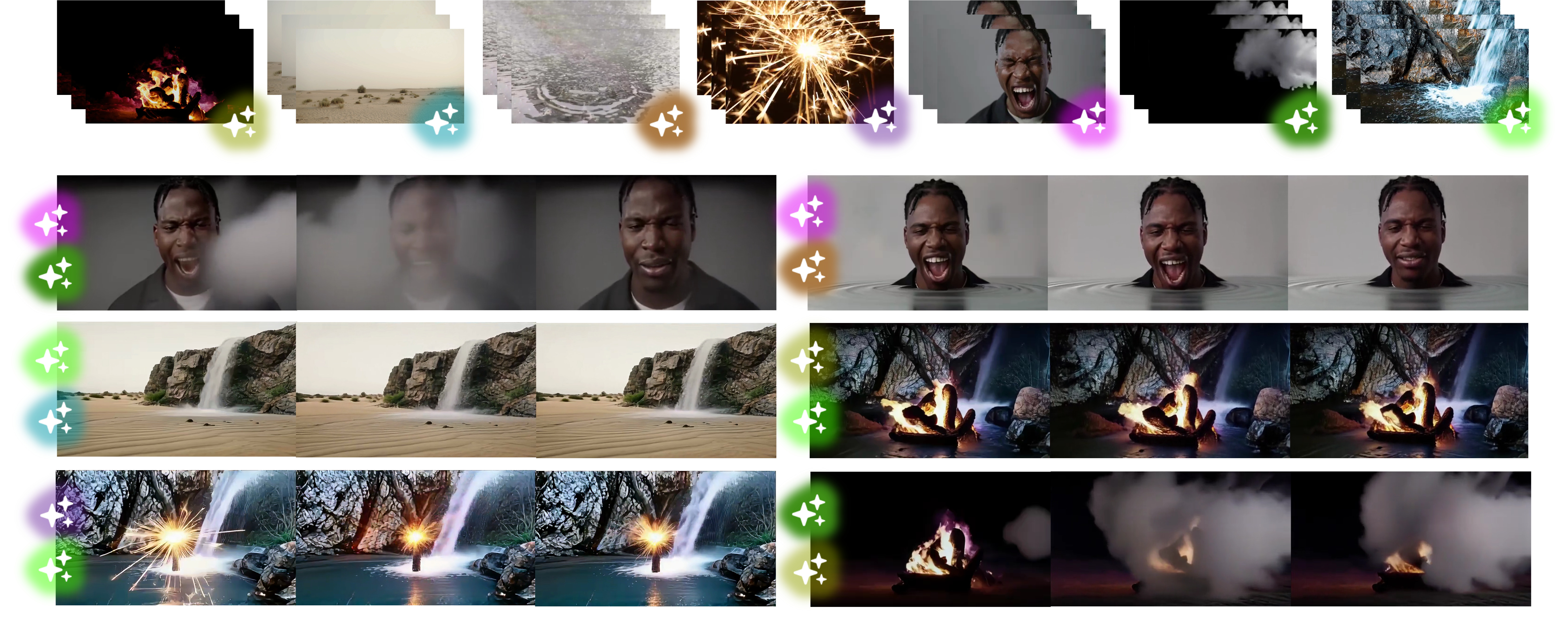}
\end{center}
\vspace{-0.3cm}
\caption{\textbf{Feed Forward Dynamic Concepts Composition.} Feedforward composition results achieved by our Grid-Fill LoRA on the composition task with each concept color-coded for clarity. For a more comprehensive demonstration, refer to the supplementary videos.}
\label{fig:composition}
\vspace{-0.3cm}
\end{figure*}

\subsection{Evaluation Metrics}  

We evaluate videos using three complementary quantitative metrics:

\textbf{Semantic Alignment (C-T).} We compute CLIP-Text Similarity~\cite{CLIP} by encoding the input prompt and each video frame, then averaging the cosine similarity between the text embedding and the mean of the frame embeddings. This captures how well the generated content matches the intended description.

\textbf{Identity Preservation (ID).} To ensure the subject’s appearance remains consistent, we use ArcFace identity similarity~\cite{ArcFace}. We extract ArcFace embeddings from the reference and generated videos and compute their cosine similarity, measuring how faithfully the individual’s identity is retained.

\textbf{Temporal Coherence (TC).} We measure frame-to-frame consistency by encoding each frame with the DINO-ViT-S/16 image encoder~\cite{caron2021emerging}, then computing the average cosine similarity between all consecutive frame pairs. Higher TC indicates smoother motion and fewer temporal artifacts.

\begin{table}[h]
\centering
\caption{\textbf{Ablation of Design Choices.} The table evaluates Identity Preservation (ID), CLIP-T (C-T), and Temporal Coherency (TC) on ablations against two design choices i.e. augmenting the training set with Grid LoRA generated samples and removing the Multi-DC LoRA during inference. }
\label{tab:ablation}
\begin{tabular}{lccc}
\toprule
\textbf{Method}       & \textbf{ID \(\uparrow\)}         & \textbf{C-T \(\uparrow\)}         & \textbf{TC \(\uparrow\)}         \\
\midrule
w/o grid samples              & 0.1878                                 &     0.1990                             &  \textbf{0.9982}                                \\
with Multi-DC LoRA           &      0.4237                    &   0.2037                       &   0.9979         \\
\textbf{Ours}  & \textbf{0.5750}               & \textbf{0.2194}                  & 0.9965                           \\
                  
\bottomrule
\end{tabular}
\end{table}

\vspace{-0.3cm}
\section{Experiment Results}

\subsection{Quantitative Evaluation}  
In this section, we evaluate dynamic concept preservation and editability across three representative editing tasks: \textit{"under purple light and smoke"}, \textit{"under rain"}, and \textit{"light illuminating the hair"}.

We begin with an ablation study examining two core design choices that enable high quality and identity consistency of our dynamic concept generation pipeline, as shown in Table~\ref{tab:ablation} and Fig.~\ref{fig:ablation}. First, we evaluate the role of Grid LoRA, whose sampled outputs are used to augment the dataset for training Grid-Fill LoRA in the later stage. Without these augmentations, the model struggles to generalize to novel scenes and edits. Second, we ablate the retention of Multi-DC LoRA. While Multi-DC LoRA provides initial inductive bias from the dataset, we observe that it often fails to reconstruct fine-grained identity details, instead generating a biased or averaged version of the subject. Removing Multi-DC LoRA and relying solely on the Grid-Fill LoRA yields improved fidelity, especially under complex edits, preserving both identity and motion.

Next, we compare our method against several competing baselines, with results presented in Table~\ref{tab:quant}. Our method consistently achieves a more favorable balance between editability and identity preservation. For instance, while DB-LoRA (Dreanbooth-LoRA) often achieves high identity similarity (ID) due to per-instance overfitting, it suffers in semantic alignment (CLIP-Text similarity, or C-T), indicating a lack of true editability. Similarly, methods like DreamVideo can exhibit high temporal coherence (TC) when motion is limited, but fail to preserve appearance under challenging edits. Qualitative comparisons in Fig.~\ref{fig:comparison} further highlight these differences. Our method is notably more robust in maintaining both appearance and motion consistency across a wide range of editing conditions. For additional qualitative results and assessments of motion fidelity, we refer the reader to the supplementary video.

\begin{table}[h]
\centering
\caption{\textbf{Editing Task Evaluation.} The table evaluates Identity Preservation (ID), CLIP-T (C-T), and Temporal Coherency (TC) on the editing task. Best scores are \textbf{bolded}, second best are \underline{underlined}. Our method achieves better ID preservation-Editing capability tradeoff.}
\label{tab:quant}
\begin{tabular}{lccc}
\toprule
\textbf{Method}       & \textbf{ID \(\uparrow\)}         & \textbf{C-T \(\uparrow\)}         & \textbf{TC \(\uparrow\)}         \\
\midrule

DreamVideo            & 0.4477                           & \underline{0.2133}               & 0.9868                           \\
NewMove               & 0.5280                           & 0.1943                           & 0.9960               \\
DreamMix              & 0.5542                           & 0.1904                           & \textbf{0.9983}                  \\
DB-LoRA               & \textbf{0.5967}                  & 0.1906                           & \underline{0.9981}               \\
\textbf{Ours}  & \underline{0.5750}               & \textbf{0.2194}                  & 0.9965                           \\
\bottomrule
\end{tabular}
\vspace{-0.5cm}
\end{table}

\subsection{Qualitative Results}  
Our framework comprises of multiple components, beginning with the sampling using Grid LoRA and followed by the personalization of dynamic concepts using Grid-Fill LoRA. We explore two primary applications in this domain: \textit{editing} and \textit{composition}.

\subsubsection{Sampling.}  
We first present results from sampling using Grid LoRA. As shown in Fig.~\ref{fig:sampling}, our model generates characters with consistent identity and motion across different grid cells. For composition, we adopt a layout where the top row contains individual inputs and the bottom row shows their compositions. As discussed in the methodology, these sampled grids serve as augmented training data for the Grid-Fill LoRA used in the personalization stage. A key ablation reveals the importance of our masked attention mechanism for the composition task, which significantly reduces cross-pane concept leakage. As illustrated in Fig.~\ref{fig:leaking}, naive application of LoRAs results in visual contamination between panes, while our masking strategy enforces spatial independence. Furthermore, our sampling framework can be extended to story generation, where the same character appears across diverse poses and environments in a consistent narrative arc (Fig.~\ref{fig:story}). These examples are provided in the supplementary video.

\subsubsection{Editing.}  
Our framework enables both \textit{global} edits and dynamic interactions, as demonstrated in Fig.~\ref{fig:editing}. The central objective here is to achieve non-pixel-aligned editing where the dynamic concept meaningfully interacts with its environment rather than being statically overlaid. We show that even though Grid-Fill LoRA is trained on a limited number of examples (25), primarily human-centric, it generalizes to a variety of out-of-domain settings. For example, we perform edits such as integrating smoke and light that dynamically interact with a person, or a cat partially submerged in water, preserving both motion and appearance. These results highlight our ability to truly personalize dynamic concepts instead of performing shallow appearance-based edits. Refer to the supplementary video for more detailed results.

\subsubsection{Composition.}  
We further support \textit{feed-forward composition} of dynamic concepts, illustrated in Fig.~\ref{fig:composition}. Here, we compose semantically diverse elements within a single coherent scene. Examples include reflective water surfaces responding to various lighting sources such as bonfires or sparklers as well as smoke blending naturally into fire, producing diffusion-like visual transitions. These compositions are not only spatially plausible but also temporally coherent, demonstrating that our method successfully captures and integrates multiple dynamic concepts in a single pass without explicit optimization or fine-tuning.
% \vspace*{-0.2cm}

\subsection{User Study}  
We conducted a user study (Table~\ref{tab:user_study}) with 10 participants, comparing ours against the baselines on adherence to prompt, motion realism, and overall preference. Participants viewed paired videos and selected the better-performing one per criterion. Here again our method shows better tradeoff, with overall performance exceeding 70\% on per-concept fine-tuning methods(DreamMix and DB-LoRA) and almost 100\% on Unet-based approaches (NewMove and DreamVideo).

\begin{table}[h]
\centering
\caption{\textbf{User Study.} User study results comparing methods on Identity Preservation (IP), Motion Preservation (MP), Adherence to Prompt (AP), and Overall Preference of the edits (OP). Preference is computed in percentages.}
\label{tab:user_study}
\begin{tabular}{lcccc}
\toprule
\textbf{Comparison}       & \textbf{IP} & \textbf{MP} & \textbf{AP} & \textbf{0P} \\
\midrule
Ours vs.\ DreamMix     &  0.32
& 0.59
& 0.98
& 0.72      \\
Ours vs.\ DB-LoRA 
  & 0.51
  & 0.58
  & 0.96
  & 0.74\\
Ours vs.\ NewMove    

  & 0.98
  & 0.95
  & 1.00
  & 1.00   \\
Ours vs.\ DreamVideo      
& 0.97
&0.99
&1.00
&0.99  \\
\bottomrule
\end{tabular}
\vspace{-0.3cm}
\end{table}

%% file: 5_conclusion.tex
\section{Limitations}

Although our approach enables fast, feed-forward personalization and composition, it has several inherent limitations. First, to balance memory and speed, we perform all LoRA inference at half the native spatial resolution and then upsample back to the original size.  Second, because our adapters operate in a single forward pass without per‐instance fine-tuning, the upper bound remains the identity preservation of per‐video LoRA methods trained directly on each example. Third, the quality of both reconstruction and downstream edits is fundamentally limited by the base video diffusion model’s capacity. Unusually aggressive or erratic motion patterns (e.g., acrobatic flips, whip-like gestures) may not be captured perfectly by the underlying DiT backbone, making such dynamic concepts difficult to personalize or compose without visible artifacts.

\section{Conclusion}

We have presented a fully feed-forward, non–pixel-aligned framework for dynamic concept personalization in text-to-video diffusion models. By training a single shared Dynamic Concept LoRA to capture entangled appearance and motion, and equipping it with lightweight, few-shot task-specific LoRA modules for editing and inpainting, our method enables zero-shot, single-pass personalization, editing, and composition of arbitrary video subjects without any per-video fine-tuning. Experiments demonstrate that our method produces temporally coherent, semantically faithful results across a wide range of subjects and effects, outperforming state-of-the-art methods. In future work, we plan to extend our approach to longer sequences and explore multi-modal conditioning for richer user control.

%% file: 6_supplementary.tex
% \clearpage

\begin{figure*}[t]
\begin{center}
\includegraphics[width=\linewidth]{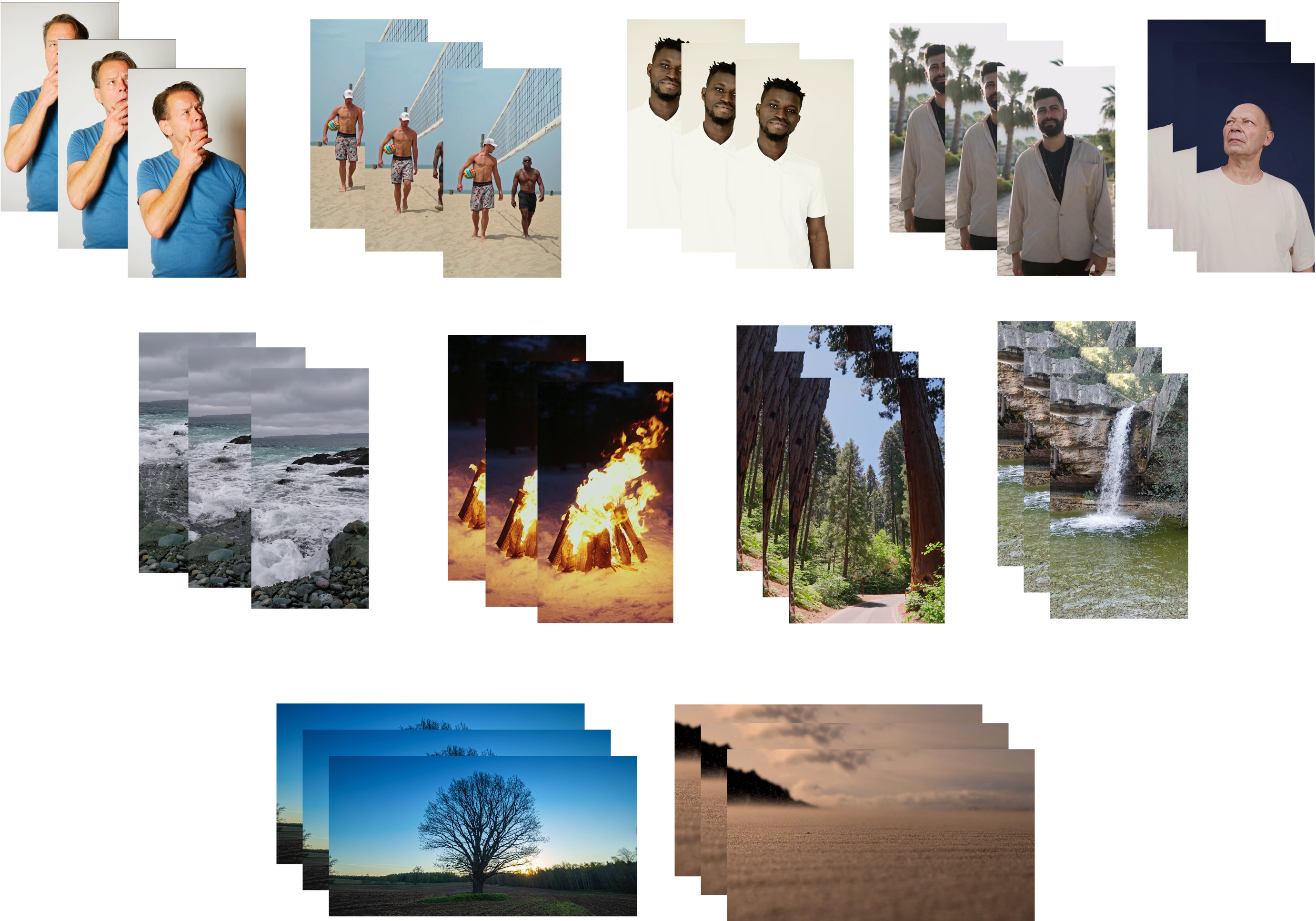}

\end{center}

\caption{Dataset used to train Multi-DC LoRA.}
\label{fig:dataset}
\end{figure*}

\section{Architecture and Training Details}

We design our system as a video diffusion model that operates in the latent space of a video autoencoder. The latent representation follows a causal video autoencoder architecture based on MAGVITv2~\cite{yu2024language}, offering a high compression ratio of $8\times16\times16$ across the temporal and spatial dimensions, and reducing the data to a compact bottleneck with 32 channels. Complete model specifications are provided in Table~\ref{tab:autoencoder_magvit}.

For the diffusion backbone, we adopt a 11.5B parameter DiT model~\cite{Peebles2023DiT}, detailed in Table~\ref{tab:backbone_dit}. This model consists of 32 DiT blocks with a 4096-dimensional hidden size. Each block comprises a self-attention layer, a cross-attention layer for integrating text inputs, and a multi-layer perceptron (MLP) with a $4\times$ expansion factor. To reduce input dimensionality, we use a $1\times2\times2$ ViT-style patchification~\cite{Dosovitskiy2021ViT}, which increases the effective compression to $8\times32\times32$. This setup enables us to process a 121-frame video at $1024\times576$ resolution using only 9216 tokens. Full 3D self-attention is employed to model high-quality motion dynamics~\cite{polyak2024movie}, with computational cost mitigated through a 6144-token attention window. Each attention block uses 32 heads and includes QK-Normalization~\cite{EsserKSD3}. Additionally, 3D-RoPE~\cite{su2024roformer} positional embeddings are applied to attention head channels in a $2:1:1$ ratio across temporal and spatial dimensions. Text prompts are encoded with T5~\cite{raffel2020t5}, and fused with video tokens using cross-attention. Diffusion timestep information is injected via modulation, following~\cite{Peebles2023DiT}.

During pretraining, we jointly train on both image and video datasets at resolutions of 512 and 1024 pixels. Videos use $16:9$ and $9:16$ aspect ratios, while images include $16:9$, $1:1$, and $9:16$. We employ progressive training over the temporal axis, gradually increasing the number of frames from 17 to 121, corresponding to 5 seconds at a fixed frame rate of 24 fps. The training uses the AdamW optimizer~\cite{loshchilov2018decoupled} with a learning rate of $1e-4$, 10k warmup steps, weight decay of 0.01, $\beta = [0.9, 0.99]$, and $\epsilon = 1e-8$, for a total of 822k steps. Training is accelerated using flash attention~\cite{dao2023flashattention2} in bf16 precision, and distributed across 256 H100 GPUs using FSDP~\cite{zhao2023pytorchfsdpexperiencesscaling}.

We follow the experimental setup of DC \textit{Set-and-Sequence}~\cite{abdal2025dynamic} to train our Multi-DC LoRA (see Table~\ref{tab:training_stages}). The Grid LoRA module is trained under two distinct configurations: composition mode and consistent dynamic concepts mode. Both modes use a dropout rate of 0.5 and a LoRA rank of 8. Training is conducted for \textit{2000} and \textit{3000} steps respectively, using the same optimization settings as Multi-DC LoRA.

For training the Grid-Fill LoRA, we apply specific modifications: we disable self-conditioning, and reduce the diffusion noise scale (sigma) to improve fine-grained control. In the consistent dynamic concepts setting, we use a sigma of 1.90 during training (1.80 at inference), while for composition mode, the sigma is set to 1.75 during training (1.70 at inference). Grid-Fill LoRA is trained for approximately \textit{3900} steps for the consistent mode and \textit{3500} steps for the composition mode. At inferernce, $cfg$ is chosen between 4 - 5 for best results.

To encourage stronger reliance on visual input, we apply text token masking with a probability of 0.6 during training. The consistent dynamic concepts mode is trained on 5 human subjects, while the composition mode includes a diverse set of 11 dynamic concepts, spanning 8 human identities and 6 non-human scenes.

\begin{table}[h!]
\centering
\begin{tabular}{@{}ll@{}}
\toprule
\textbf{Autoencoder}      & \textbf{MAGVIT}                        \\ \midrule
Base channels             & 16                                     \\
Channel multiplier        & [1, 4, 16, 32, 64]                     \\
Encoder blocks count      & [1, 1, 2, 8, 8]                        \\
Decoder blocks count      & [4, 4, 4, 4, 4]                        \\
Stride of frame           & [1, 2, 2, 2, 1]                        \\
Stride of h and w         & [2, 2, 2, 2, 1]                        \\
Padding mode              & replicate                              \\
Compression rate          & \(8 \times 16 \times 16\)              \\
Bottleneck channels       & 32                                     \\
Use KL divergence         & \checkmark                             \\
Use adaptive norm         & \checkmark (decoder only)              \\ \bottomrule
\end{tabular}
\caption{Autoencoder and MAGVIT specifications.}
\label{tab:autoencoder_magvit}
\end{table}

\begin{table}[h!]
\centering
\begin{tabular}{@{}ll@{}}
\toprule
\textbf{Backbone}               & \textbf{DiT}                     \\ \midrule
Input channels                  & 32                               \\
Patch size                      & \(1 \times 2 \times 2\)          \\
Latent token channels           & 4096                             \\
Positional embeddings           & 3D-RoPE                          \\
DiT blocks count                & 32                               \\
Attention heads count           & 32                              \\
Window size                     & 6144 (center)                    \\
Normalization                   & Layer normalization              \\
Use flash attention             & \checkmark                       \\
Use QK-normalization            & \checkmark                       \\
Use self conditioning           & \checkmark                       \\
Self conditioning prob.         & 0.9                              \\
Context channels                & 1024                             \\ \bottomrule
\end{tabular}
\caption{Backbone and DiT specifications.}
\label{tab:backbone_dit}
\end{table}

\begin{table}[b]
\centering
\begin{tabular}{@{}lcc@{}}
\toprule

Optimizer                   & \multicolumn{2}{c}{AdamW}              \\
Learning rate               & \multicolumn{2}{c}{$1 \times 10^{-4}$} \\
LR scheduler                & \multicolumn{2}{c}{constant}           \\
Beta                        & \multicolumn{2}{c}{[0.9, 0.99]}        \\
Weight decay                & \multicolumn{2}{c}{0.01}               \\
Gradient clipping           & \multicolumn{2}{c}{0.05}  
                \\\bottomrule
\end{tabular}
\caption{Optimization settings.}
\label{tab:training_stages}
\end{table}

\section{Prompts}

Providing rich, structured prompts at initialization is key to steer our feed‐forward LoRA adapters towards precise video editing and composition. Each prompt begins with a \emph{global descriptor} that succinctly names the subject and its primary motion (e.g., “A cat [Cat\_identity] performing a leisurely stroll [motion\_stroll]”), ensuring the model anchors on the correct dynamic concept. For editing tasks, we append \emph{region‐specific modifiers} that specify per‐pane variations such as background, lighting, or weather while retaining the same identity and motion tags, so that each quadrant reflects the \emph{exact same} underlying appearance and motion under different contextual conditions. For composition tasks, prompts enumerate the two source concepts in separate “original input” clauses (e.g., fire and a dancing figure), followed by parallel “compositional result” clauses that describe how the identical subject motion and appearance are merged with the identical secondary concept. This structured, multi‐clause prompt design—combining global descriptors, modifier tags, and explicit composition labels—enables our specialized LoRA modules to faithfully reproduce non–pixel‐aligned edits and seamless concept fusions in a single forward pass.

\medskip

\noindent\textbf{Example (Editing Prompt):}
\begin{quote}
This 2×2 grid illustrates consistent generation of a cat [Cat\_identity] performing the exact same leisurely stroll [motion\_stroll]. [TOP LEFT] Cat [Cat\_identity] performing the exact same leisurely stroll [motion\_stroll] on a blue playground surface. [TOP RIGHT] Cat [Cat\_identity] performing the exact same leisurely stroll [motion\_stroll] submerged in clear water up to its chest. [BOTTOM LEFT] Cat [Cat\_identity] performing the exact same leisurely stroll [motion\_stroll] in heavy rain, its fur drenched. [BOTTOM RIGHT] Cat [Cat\_identity] performing the exact same leisurely stroll [motion\_stroll] at night under violet ambient lighting.
\end{quote}

\noindent\textbf{Example (Composition Prompt):}
\begin{quote}
This 2×2 grid of videos illustrates the creative fusion of Lincoln [Lincoln\_identity] performing the exact same angry opening gesture then composing himself to look directly at the camera [motion\_angry\_gesture\_look] and a dancing spark effect [SparkEffect]. [TOP LEFT] shows the original input for Lincoln: Lincoln, a young Black man with short braided hair and wearing a dark shirt, performing the exact same angry opening gesture then composing himself to look directly at the camera [motion\_angry\_gesture\_look] against a plain grey studio background. [TOP RIGHT] shows the original input for the effect: Golden sparks flying outward from a central point against a dark backdrop, capturing the exact same dynamic spark motion [SparkEffect]. [BOTTOM LEFT] presents a compositional result: Lincoln [Lincoln\_identity] performing the exact same gesture [motion\_angry\_gesture\_look] surrounded by the exact same dancing sparks [SparkEffect], the sparks reflecting off his braided hair and intense expression. [BOTTOM RIGHT] presents a second compositional result: Lincoln [Lincoln\_identity] performing the exact same gesture [motion\_angry\_gesture\_look] with the exact same golden sparks [SparkEffect] weaving around his body, creating a dramatic, high‐energy scene.
\end{quote}

\section{Video Sources}
The videos used in this work are taken from \href{https://www.pexels.com/}{Pexels}. Fig.~\ref{fig:dataset} shows the videos on which Multi-DC LoRA was trained on. These samples are taken from set-and-sequence~\cite{abdal2025dynamic}.

\section{Impact Statement}

Generative AI technologies, including the personalized video generation techniques explored in this paper, offer significant potential for creative expression, storytelling, education, and accessibility applications. We are committed to the responsible advancement of generative AI and encourage its application towards beneficial goals, such as enhancing creative tools, improving virtual production pipelines, and creating novel accessibility features.